# Enjeux normatifs des TICE de l'enseignement des langues dans le contexte arabo-berbère

**Henri Hudrisier,**
Laboratoire Paragraphe, Université de paris 8, France

**Mokhtar Ben Henda,**
CEMIC-GRESIC, Université Bordeaux 3, France

*Résumé :*

*L'e-Learning est maintenant un phénomène mondial. Apprendre l'arabe (ou des dialectes arabes) ou apprendre une ou plusieurs variantes du berbère peut s'entendre dans un contexte très local (au Maghreb) ou plus largement dans un contexte de la diaspora ou plus largement encore dans un cadre mondial (un japonais ou un russe qui apprend l'arabe ou le berbère). Les ressources d'enseignement à distance ainsi crées doivent être utilisables potentiellement dans n'importe quel contexte culturel et linguistique du monde. Cela implique de ce fait que les ressources ainsi créées puissent d'abord répondre au cadre général de normalisation qui se met en place au niveau de l'ISO/IEC JTC1 SC36, et pour bien des aspects au-delà de cette seule instance de normalisation.*

*Mots-clés: TICE, normalisation, multilinguisme, apprentissage des langues, langue arabe, langue berbère*

*Abstract :*

*E-learning is becoming a global phenomenon. Learning Arabic (or Arabic dialects), or learning one or several variants of Berber can be understood from a very local perspective (in the Maghreb for instance) or in the wider framework of the diaspora or even more broadly in a global world context (in case a Japanese or a Russian learns Arabic and Berber). Resources for distance learning must then be created and potentially used in any international cultural and linguistic context. This implies that the resources created for such perspective should cope with the general standards framework of the ISO / IEC JTC1 SC36, and even beyond the scope of this standardization instance.*

**Introduction** :

La mondialisation est un état de fait. Avec la technologie numérique globalisée ce sont sans doute deux facettes très déterminantes de l'histoire du début du 21ème siècle. La mondialisation tient à de nombreuses réalités concomitantes entre autres : l'explosion démographique et le progrès technologique notamment celui des techniques de communication tant pour le transport des personnes que l'intercommunication globale des réseaux d'informations. Cela confronte la totalité des cultures et des langues à l'urgence d'inventer et de développer les moyens techniques qui faciliteront une fluidité interculturelle et interlinguistique. Le génie traductique et l'apprentissage des langues notamment avec les TIC en sont les 2 piliers.

**La normalisation des TIC et les TICE**

Dans cette contribution nous ne nous focaliserons pas, contrairement sans doute à la majorité des intervenants, à l'étude d'un dispositif ou d'une méthode particulière d'apprentissage des langues dans le contexte arabo-berbère. Notre ambition consistera à faire un survol de l'état de l'art normatif des TIC qui constituent l'environnement de la communication et de l'information linguistique, de la codification des écritures et des TICE. Nous confronterons cet état de l'art au contexte linguistique, culturel et territorial du monde arabo-berbère.

Nous savons que cette contribution techno-normative peut apparaître comme marginale mais nous sommes cependant persuadés qu'elle est, non seulement utile, mais d'année en année plus indispensable pour tous les acteurs de l'enseignement médiatisé par les TICE et plus nécessaire encore pour ceux de l'enseignement des langues.

Certes, le chantier de la normalisation des systèmes d'information est relativement récent mais la mondialisation galopante que nous mettions en avant au tout début de cet article rend de plus en plus indispensable et incontournable une prise en compte approfondie des normes et standards. Les informations qui circulent partout autour du monde ne peuvent, en effet, le faire que parce qu'elles répondent à des règles de conformité de formats et d'interopérabilité des dispositifs. Pour les informations textuelles, il est indispensable qu'elles soient conformes d'une part à une norme unique et universelle de représentation des caractères (ISO/IEC10646 et le standard Unicode correspondant), qu'elles correspondent à des normes morpholexicales ou morphosyntaxiques, et d'autre part, pour les informations audiovisuelles et multimédia (notion très importante dans l'apprentissage des langues tant par images que par sons), il faudra prendre en compte les normes du domaine (notamment celle de l'ISO/IEC JTC1 SC29, familles de normes JPEG et MPEG) mais aussi de nombreuses normes et standards propres à la synthèse mais aussi à la reconnaissance vocale.

Notons aussi que le Web est aussi devenu une composante consubstantielle de tout système informatique et dans moins de 5 ans le réseau numérique intégral se généralisera intégrant non seulement l'informatique, la téléphonie mobile (et bien sûr fixe) comme c'est déjà le cas, mais aussi la TV et la radio, étant entendu que tout *broadcast* analogique sera définitivement abandonné au tournant des années 2010. Or, dans la culture des jeunes, mais aussi des moins jeunes, l'accès à un service ou à un produit passe presque nécessairement par le réseau. Même si de nombreux pans de l'offre des contenus d'enseignement électroniques est encore hors ligne, il devient de plus en plus exceptionnel qu'une ressource pédagogique puisse exister hors du Web et a fortiori qu'elle ne soit pas au minimum référencée dans une offre en ligne. Là aussi il y a matière à normalisation indispensable pour rentabiliser la création de contenus d'enseignement médiatisé par des TICE : le CDM, (*Courses Description Metadata*), une norme qui est en cours de généralisation au niveau européen au CEN, (*Comité Européen de Normalisation*) sous le nom de MLO, (*Metadata for Learning Opportunities*).

Ceux qui pensent qu'il est trop tôt pour aborder des questions normatives qui concernent plus le devenir des TICE que ses applications actuelles concrètes, argueront que le contexte Maghrébin est différent. Que les pays du Nord sont plus avancés que les pays du Sud. Cela reste à voir, car presque partout dans le monde, des rives du Mékong aux oasis sahariennes, le Web café n'est jamais bien loin (Boase et al., 2003). Mais cela deviendra de plus en plus vrai à partir du moment (tournant des 2015) où la diffusion radio, TV, téléphone et informatique ne seront plus qu'un seul et même réseau numérique intégré : un web.2++ globalisé et convergeant. Il faut en effet noter que les pays du Sud ont sans nul doute plus d'urgence (et sans doute d'opportunités) à adopter rapidement les technologies numériques les plus avancées à partir du moment (qui n'est pas lointain) où celles-ci se banalisent sur des réseaux déjà largement diffusés à bas coût partout dans le monde (c'est le cas de la radio, de la TV et du téléphone portable), et à partir du moment où même l'informatique personnelle devient disponible pour des coût qui deviennent inférieur à 100$ avec une autonomie par rapport au réseau câblé de plus en plus performante et une faible consommation électrique remarquable (qui se conjugue avec la possibilité de disposer à bas coût de batterie solaires). On remarquera aussi que même si les Etats du Sud sont peu représentés dans les instances de normalisation des TICE (à l'exception remarquable de l'Algérie et du Kenya), cela ne signifie pas pour autant qu'ils n'ont pas (plus qu'au Nord) une obligation de réussir au plus tôt la normalisation de leurs ressources et de leur réseau matériel de TICE. En effet la faible taille relative du parc d'équipement tant de réseaux, que de plates-formes et des fonds de ressources pédagogiques leur permettra paradoxalement de parvenir sans difficultés à accepter des normes, là où les pays développés subissent l'handicap d'avoir à assurer la mise à niveau des anciennes TICE non normalisées et la difficulté d'avoir à convaincre leur communauté académique (et celle des gestionnaires) de changer leurs matériels existant, de modifier leurs habitudes de gestion des ressources pédagogiques et de procéder à la reprise historique de fonds de contenus quelquefois considérables. Il est à noter aussi que la rentabilité de l'ingénierie linguistique étant plus faible en arabe qu'en anglais, et aujourd'hui insignifiante en langue berbère, les experts du domaine ne peuvent compter que sur eux mêmes et ils ne parviendront à développer des outils que en s'alliant en synergie dans un cadre qui sera obligatoirement consensuel et normatif.

**Des enjeux techno-normatifs paradoxaux selon le contexte économique, culturel, linguistique ou géographique :**

La norme de codification des écritures[1] (et le standard Unicode qui lui correspond) est un excellent exemple du retournement paradoxal des enjeux techno-normatifs selon le contexte linguistique. Jusqu'à la fin des années 70' il était très difficile de faire de l'informatique autrement qu'en anglais ou à la limite dans des écritures alphabétiques (si possible latines) s'écrivant de gauche à droite. C'est précisément au début des années 80' que les Japonais (et dans leur sillage les Chinois) ont décidé de développer une codification des caractères idéographiques correspondant à leur besoins jusqu'ici impossibles à satisfaire. Il faut rappeler que ces nouveaux développements étaient devenus possibles grâce à des progrès techniques informatiques qui émergeaient à cette époque[2]. Pour ce faire les industriels et les chercheurs du Sud Est asiatique ont dû bousculer la vieille norme ASCII qui était devenu un des fondements de l'informatique traditionnelle. Ils ont du proposer de changer de paradigme et de se baser sur une informatique à 16 bits et non plus 8 bits et ce faisant ils se sont trouvés en situation de proposer une offre potentiellement omni-écriture. En effet, en faisant évoluer la « casse typographique virtuelle » de 8 à 16 bits[3] (André & Hudrisier, 2002), il devenait dès lors possible de coder, avec la combinaison de 2 octets, non seulement l'écriture chinoise, japonaise, coréenne mais aussi toutes les grandes familles d'écritures alphabétiques européennes, l'arabe, l'hébreu et les écritures indiennes.

---

[1] ISO/IEC10646
[2] Capacité pour les ordinateurs d'évoluer du 8 bits vers le 16 et même 32bits.

Dans le monde des experts en normalisation ce retournement de conjoncture est un moment charnière dans l'histoire techno-normative[4]. Les industriels et les chercheurs informatiques du reste du monde (non idéographique) ont du accélérer leur démarche et développer en synergie avec les premiers standards de code caractère japonais la solution à 32 bits (4 octets) qui correspond aujourd'hui à Unicode. Cela prouve bien qu'une situation de domination culturelle, technologique, linguistique, géographique n'est jamais définitivement acquise.

Notre hypothèse est que le Maghreb, et plus encore la communauté franco-arabo-berbère est en situation d'handicaps, ou d'avantages, selon la lecture qu'on veut en faire qu'il lui appartient de transformer en prospérité technolinguistique.

**Avantage multilingue du Maghreb arabo-berbère**

Un berbère instruit (et la société maghrébine ne manque pas de jeune diplômés) est au minimum trilingue à l'oral[5], et avec un peu d'effort il peut donner une grande valeur ajoutée à ce potentiel multilingue premier s'il sait déployer ce premier niveau de compétences sur un savoir linguistique écrit. Il disposera alors de la maitrise de l'arabe littéraire, d'une compétence pour écrire le berbère, en distinguer au minimum une forme standard de sa langue berbère maternelle ou mieux une compétence entre différents berbères et un savoir de transcription dans les 3 alphabets (arabe, tifinagh, latin). Il entretiendra par ailleurs sa compétence francophone tant à l'oral qu'à l'écrit ainsi bien sûr que sa connaissance d'autres langues (notamment l'anglais). Cet avantage multilingue et multiécriture est un atout fondamental pour la société monde qui recherche (souvent sans succès) des personnes multilingues (et surtout multilingues entre des langues appartenant à des familles d'écritures différentes). Or chacun sait que l'obstacle de l'apprentissage d'une nouvelle langue (ou d'une nouvelle écriture) se franchit d'autant mieux que l'apprenant est déjà bi ou trilingue (mais aussi encore bi ou tri écriture).

En dehors des considérations d'identité culturelle et linguistique, il y a donc là un argument économique supplémentaire : maitriser 3 langues et 3 écritures est un fabuleux avantage pour apprendre d'autres langues, et notamment pour apprendre les langues non européennes qui deviennent de plus en plus indispensables pour être un acteur efficient dans la mondialisation. Pour cela, et contrairement à ce que l'on peut souvent constater chez certains de nos étudiants (notamment ceux qui sont des magrébins vivant en France depuis 1 ou 2 générations) il n'est pas inutile de maitriser le français ET l'arabe ET le berbère à l'écrit comme à l'oral; pour l'arabe, l'arabe classique EN PLUS de l'arabe dialectal et pour le berbère, sous forme orale ET écrite (en connaissant une forme standard de son berbère d'origine avec si possible ses 3 modes d'écritures et en ayant une connaissance au moins générale de l'aire linguistique berbère).

---

[3] La codification des caractères, dans son principe, peut se ramener à la création d'une « casse typographique virtuelle ». Celle-ci présente autant de cases élémentaires permettant de ranger (de coder) un caractère qu'il est possible d'avoir de positions possibles (de codages possibles) générés par la combinaisons des bits d'un octet (256) de 2 octets (65536) ou de 4 octets (plus de 4 millions)
[4] Un moment aussi important de l'histoire techno numérique que le sera le tout numérique audiovisuel donc multimédia et global induit par l'abandon du broadcast analogique au tournant des années 2012.
[5] Un magrébin non berbère instruit est au minimum bilingue et maitrise deux écritures.

Développer cet avantage du multilinguisme maghrébin comme un atout pour affronter la prospérité dans la mondialisation passe bien sûr par la mise à disposition en réseaux de ressources pédagogiques d'apprentissage des 4 langues fondamentales pour le Maghreb : arabe, berbère(s), français et anglais. Cela passe aussi par le développement d'une ingénierie linguistique dans un contexte de faible retour sur investissement et d'une appropriation technique par les utilisateurs et surtout les experts linguistes magrébins seuls à même de participer au développement de ces outils dans un contexte de logiciel libre.

**S'inscrire dans une démarche participative de la normalisation des ressources linguistiques mais aussi celle des TIC et des TICE**

Qui dit enseignement des langues suppose qu'on puisse disposer de ressources linguistiques dans les langues que l'on veut enseigner. Suppose aussi que l'on dispose d'environnements technos logiciels, de plates-formes adaptées, tant au traitement linguistique (terminologie, lexicographie, alignement parallèle de corpus, traitement de l'écrit, traitement de l'oral) qu'à la maitrise multimédia des fonctionnalités d'un labo de langue[6] et enfin des outils ou plates-formes adaptées à la pédagogie et plus encore à la pédagogie spécifique d'apprentissage des langues.

Les normes sont malheureusement trop souvent considérées comme une contrainte externe (dont on doit tenir compte quand on leur est confronté) alors que de nombreuses normes sont en fait des « auberges espagnoles ». Elles ne permettent de gérer telle ou telle spécificité culturelle ou technique qu'autant que l'on se sera préoccupé d'implémenter ce potentiel en amont (et si ça ne l'était pas que l'on se sera mobilisé pour que cela devienne possible).

C'est précisément ce qu'ont réalisé les experts de l'IRCAM. Ils voulaient disposer de l'écriture tifinagh. Ils ont constaté qu'elle était absente des premières versions d'UNICODE et de la norme ISO/IEC10646 et ils ont fait en sorte qu'elle soit proposée puis acceptée. Aujourd'hui d'autres berbères peuvent penser (avec raison) que certains caractères manquent. Il y a 2 attitudes :

- Ils peuvent penser que la norme est inadaptée et que l'on n'a pas envisagé toutes les variantes possibles du tifinagh.
- Ils devraient penser (c'est plus positif) que l'espace réservé pour l'écriture tifinagh dans la norme prévoit des espaces laissés libres pour coder d'autres caractères. Il y a donc urgence pour qu'ils constituent une commission d'experts pour « réviser UNICODE et la norme ISO/IEC10646 » et y apporter des additifs dans la zone berbère.

Il serait opportun que ce comité d'écriture pan-berbère examine aussi les questions touchant à la codification Unicode non seulement des variantes modernes du tifinagh mais aussi des écritures archéologiques mais aussi encore des questions éventuellement posées par la transcription latine et arabe (il ne devrait pas y en avoir). Il leur faudra aussi s'enquérir de l'avancement des travaux pour la réalisation graphique et logicielle des polices contenant les caractères berbères effectivement disponibles dans des environnements banalisés (explorateurs web, éditeurs, parseurs, etc...).

Toutes ces ressources d'écritures et leurs variantes supposent (pour pouvoir être référencées correctement) que l'on dispose de codes de langues et de codes d'écritures normalisés exhaustifs, non ambigües en distinguant clairement les codes et variantes d'écritures des codes et variantes de langues.

---

[6] Et maintenant d'outils « nomades » adaptés à l'apprentissage pervasif des langues mais aussi adaptés à un territoire faible en infrastructures tant électrique que réseau fixe de télécommunication.

**Les normes et standards d'ingénierie linguistique générale et spécifiques**

L'enseignement des langues requiert obligatoirement des outils linguistiques (et leur normes et standards hors ceux de l'écriture) de caractère général (propres à toute les langues) et/ou spécifique (propre à telle ou telle langue). La question est évidemment notablement différente s'il s'agit de l'arabe et du berbère. Autant il existe des terminologies, des dictionnaires électroniques, des outils de traduction automatiques en arabe autant il n'existe presque rien d'équivalent en langue berbère. Les experts berbères sont trop peu nombreux et disposent de trop peu de crédits de recherche et de capacités de développement préindustriel, et ils doivent de ce fait travailler avec le maximum de synergies collaboratives. Ils doivent être aussi certains que leurs travaux ne seront pas dupliqués plusieurs fois. Il importe aussi pour cela qu'ils soient ouverts sur leurs propres diversités, les synergies entre les diverses langues berbères et les variétés de parlers tant au Maroc qu'en Algérie. Ils doivent prendre en compte les équivalences de transcription dans les 3 écritures + l'API. C'est à ce prix seulement qu'on peut avoir l'espoir de voir se développer de tels outils.

La question des outils syntaxiques (grammaires électroniques) est aussi hautement stratégique. Tant pour la représentation du lexique, de la syntaxe que du sens il est très utile de se référer aux normes les plus modernes notamment la norme produite en consortium par la TEI et le TC37 dite « norme de description des caractéristiques de traits ».

**Des normes en convergence pour le multimédia, les réseaux, le sémantique et les TICE**

Dans ce futur à court terme, la normalisation des ressources pédagogiques sera devenue une obligation incontournable. Ne pas l'être, correspondra à une impossibilité d'être échangée en réseau, de pouvoir être réutilisé mais plus encore à une impossibilité d'être visualisé, écouté et utilisé sur tous types de terminaux.
Les plate formes[7] et le multimédia également sous l'influence de la normalisation convergente se banalisent, deviennent interopérable et omni-usages ce qui veut dire que tout enseignement médiatisé sera indifféremment transférable de l'ordinateur au téléphone (surtout mobile et PDA), au poste de TV, à la console de jeux et au poste de radio (éventuellement baladeur).

On voit bien dès lors les questions qui doivent être approfondies. Il ne s'agit pas seulement de la normalisation des ressources mais aussi de la normalisation, elle aussi convergente, des plateformes et autres dispositifs matériels qui permettent de créer puis d'utiliser des ressources d'enseignement des langues. On sait tous en effet que les spécialistes des labos de langues[8] sont tous des pionniers des technologies multi et hyper média. Le problème qui se posera à court terme à tous les créateurs de dispositifs technologiques favorisant l'apprentissage des langues consistera à savoir conserver l'avance conceptuelle que constitue ces dispositifs multimédia aujourd'hui encore disparates et d'en intégrer les fonctionnalités innovantes, voire spécifiques, dans la convergence normalisée du multimédia. On est là encore dans une situation « d'avantage ou d'inconvénient » selon la lecture qu'on choisit d'en faire, similaire à celle des informaticiens d'écriture latine face aux informaticiens d'écriture idéographique. Ce retournement concerne maintenant les « e-learnistes » de l'enseignement des langues en situation de handicap : eut égard à la complexité multimédia qu'induisent l'apprentissage des langues comparé à d'autres apprentissages traditionnellement plus mono-média[9].

---

[7] Plate-forme est le terme consacré pour exprimer aujourd'hui un environnement matériel et/ou logiciel spécifique pour réaliser une action (ici de l'enseignement des langues).
[8] Et de quantités d'autres dispositifs textuel et/ou audio et/ou visuel plus ou moins interactif.
[9] Evidemment, rares sont les disciplines dont l'apprentissage pourrait être qualifié de strictement mono-média. Presque toutes les disciplines, activités ou métiers peuvent trouver avantage à utiliser

Pour les pédagogues des langues intéressés par cette dimension multimédia, il devient fondamental de pouvoir réfléchir de façon globale et plus uniquement strictement fonctionnelle. Mobile-learning (ou M-learning), T-learning (pour Digital Television-learning), enhable learning ou l'EAD pervasifs sont des concepts qui sans être aucunement synonymes montrent combien la notion de convergence multimédia devient une réalité inéluctable. Même si les formats d'écrans, même si les interfaces de saisie ou de commande, même si la mobilité de tous ces systèmes est éminemment distincte, on assiste à une tendance irrésistible à ce que ces différentes plates-formes puissent utiliser des ressources communes. Une ressource d'enseignement des langues conçue pour être affichée sur un écran de taille confortable et écouté avec des écouteurs très performants dans un laboratoire de langue doit pouvoir se décliner sur la fonction « baladeur » d'un téléphone portable ou être potentiellement visionnée dans une émission de TV ou écoutée dans un programme de radio numérique pédagogique[10].

Ceci exige que les concepteurs de ressources comprennent cette tendance d'évolution technique et intègre les normes en cours de développement (notamment au SC36 mais aussi SC29) qui permettent si on les respecte de répondre naturellement à ces potentiels d'adaptation, à ces différents formats et systèmes d'information.

Dernier point, linguistique celui-ci. On assiste actuellement à de très importants chantiers de développement des normes et standards dans le domaine de la traductique et de la gestion des corpus linguistiques tant oraux que textuels (TEI, TC37 normes de description des caractéristiques de traits, etc.).

Ces normes et standards sont à considérer à trois niveaux :

1. Celui de l'enseignement spécialisé des langues pour les traducteurs et interprètes qui auront de plus en plus nécessité de connaître et savoir interagir avec des outils d'assistance à la traduction fortement dépendant des normes.
2. Il en sera indirectement de même pour les utilisateurs lambda utilisant des outils traductiques moins sophistiqués
3. Celui des professeurs de langues qui devront progressivement s'approprier les nouvelles possibilités techniques rendues possibles par des métadescriptions morphosyntaxiques des langues mais aussi des descriptions sémantiques interlinguistiques.

Les experts linguistes arabo-berbères doivent là encore se mobiliser pour que leurs langues soient aussi documentées (et par qui d'autres qu'eux ?) pour ce qui est des normes d'interopérabilité entre langues.

**La norme comme méthode mais aussi comme domaine d'intervention**

Nous avons précisé dans l'introduction de notre intervention que nous n'étudions pas une situation ou un dispositif particulier pour apprendre les langues dans le contexte arabo-berbère, mais que nous participons (très en amont) à l'aménagement du dispositif informationnel dans sa dimension normative. Nous affirmons que le respect des normes

---

plusieurs modalités de médiation (texte, image, son, aspect ludique, fonction collaborative, mobilité). On remarquera cependant que certaines disciplines (où branche de discipline) peuvent souvent se contenter de la médiation strictement textuelle (histoire, informatique, sociologie, économie…) alors que d'autres se passent difficilement du multimédia (comment faire de l'histoire de l'art sans image, comment apprendre des langues sans texte ET audio, quoiqu'il est avéré que certaines méthodes de langue se concentrent sur la seule dimension audio existantes ou l'anticipation des normes à venir est un des impératifs de méthode pour développer, gérer, diffuser des contenus d'enseignement des langues. Nous pensons que cela est d'autant plus indispensable parce que l'on peut constater que où que se situe la focalisation d'effort de recherche ou de développement d'une ingénierie de l'apprentissage des langues, elle rencontrera nécessairement plusieurs des domaines définis par des normes ou en cours de définition normatives :

- technologies et méthodes de la pédagogie : ISO/IEC JTC1 SC36.
- technologies et méthodes de la représentation des caractères et écritures :ISO/IEC JTC1 SC2, norme ISO/IEC10646 plus connu sous le standards de UNICODE.
- Normes de l'audiovisuel et du multimédia : ISO/IEC JTC1 SC29 (notamment la famille de normes MPEG, mais aussi les nombreuses normes et standards de la codification audio) ou encore SC24 (infographie)
- ISO TC37 Terminologie et autres ressources linguistiques : l'instance normative qui permet de normaliser des terminologies et des lexiques (notamment multilingues). Cette même instance développe entres autres des normes pour la gestion des ressources linguistiques (ISO24610), notamment « la représentation des structures de traits (ISO24610 -1).

[10] La TV et radio numérique étant potentiellement beaucoup plus abondante : le *broadcast* numérique multiplie par un facteur 10 la capacité de diffusion à « plage de fréquence égale ». D'autre part les émissions hertziennes numériques sont potentiellement interactives et multimédia.

On devrait aussi s'intéresser à des normes de l'UIT ou à certains standards du W3C. Mais ceci dépasse les limites d'un tel article.

**Pour une acception non polémique du concept de norme :**

Lorsque l'on aborde la question de la normalisation avec des enseignants (mais aussi avec quantité d'autres publics), nombre d'entre eux ont trop souvent tendance à considérer ce terme dans son sens trivial. Normaliser signifie pour eux égaliser, utiliser partout et pour tous selon les mêmes méthodes, les mêmes formats, les mêmes outils.

Il importe d'éclaircir cette question dans notre exposé de méthodes. Normaliser des ressources pédagogiques ou normaliser les plates formes c'est précisément le contraire du sens trivial du mot « normaliser » : cela consiste à partir de la définition en consensus de certains grands principes d'interopérabilité et de compatibilité (qui sont soit des normes, soit des standards, soit encore des bonnes pratiques d'utilisation), d'ouvrir à l'infini la totalité des possibles dans l'échange (éventuellement interlinguistique) d'une extrême diversité de contenus selon des styles pédagogiques, des modalités fonctionnelles, des types d'environnements techniques et pédagogiques aussi disparates et diversifiés qu'il est possible.

Pour comprendre ce que vise la « normalisation des TICE », observons un domaine similaire : celui des traitements de textes. Il y a une trentaine d'années si on saisissait un texte dans un environnement informatique X, il était pratiquement impossible de récupérer ce texte dans un environnement informatique Y. Tout au plus pouvait-on dans le meilleur des cas, récupérer le contenu principal de ce texte « au kilomètre » en ayant perdu toute mise en page, la structure des notes et des titres et en étant certain qu'au-delà de l'écriture latine stricto sensu, tout caractère accentué, toute cédille ou autre diacritique était irrécupérable.

La normalisation a bel et bien ouvert vers une infinité de possibles le domaine des traitements de textes en autorisant l'ouverture à toutes les langues du monde, en permettant de conserver, quel que soit le contexte, des fonctions de plus en plus complexes : les notes, la mise en page, le mode plan, le mode correction, l'adjonction de nombreux objets (images, tableaux, commentaires) et permettant que se développe et puisse s'intégrer de façon modulaire des outils (correcteurs orthographiques, outils d'aide à l'écriture, éditeurs logiciels, explorateur web, etc…)
Tout ceci n'aurait jamais été possible sans normes ou standards.

Ce dont il est d'abord question en normalisant les TICE est précisément du même ordre. Depuis déjà de nombreuses années la plupart des spécialistes de l'e-learning sont déjà persuadées de l'utilité des normes permettant la description de ressources pédagogiques comme avec le LOM[11] pour les diffuser et y accéder facilement permettant ainsi leur réusabilité. Nous dirions que c'est le niveau *basique* de la normalisation des TICE : celui qui correspondrait dans notre exemple précédent à la récupération de textes dans la diversité des écritures alphabétiques (grec, cyrillique, arabe, hébreux, hindi et bien sûr des diverses latines accentuées).

Au-delà de ce premier niveau, tout pédagogue comprendra qu'à l'instar des fonctionnalités d'un traitement de texte la pédagogie médiatisés sur des médias électroniques répond à des styles, à des fonctionnalités, à des types d'organisation du rapport entre les acteurs, à des modalités de médiation et d'interface, à des modes d'organisation institutionnelle ou de consommation d'*eductainment*, à des types de *procurement*, de gestion des droits et copyrights, à des modes d'évaluation (voire de non évaluation ou d'autres modalité de *feed-back* de l'évaluation). A l'évidence, il est indispensable de pouvoir distinguer des moments diversifiés de l'apprentissage d'une langue étrangère : proposition d'un thème, d'une version, d'une période d'apprentissage de vocabulaire (de conjugaison, de déclinaisons…), d'acquisition de grammaire (de grammaires comparées), de diction, d'écoute-compréhension, d'étude littéraire étrangères et comparées, des moments d'exercices, des flux de correction, des interactions entre professeurs, tuteurs et apprenants, etc..… Ce sont toutes ces fonctions qu'il faut définir en consensus pour en permettre la mise en œuvre aussi diversifié que possible. C'est à ce chantier normatif que répond le SC36. On comprend dès lors que ce type de normes, quand elles seront finalisées, ouvrira considérablement la créativité des e-pédagogues. Certes il existe aujourd'hui un certain nombre de plateformes qui permettent de réaliser à la demande un grand nombre de ces fonctions. Mais tout le problème est précisément que d'une plate forme à l'autre tout le travail est à recommencer. L'entreprise A qui a mis en place toute sa formation (notamment sa formation en langue), dans l'environnement d'une plate forme X ne peut pas cumuler son patrimoine de ressources pédagogiques avec celles de l'entreprise B à laquelle elle s'associe si celle-ci a travaillé sur la plateforme Y. Les éditeurs de contenus d'enseignement des langues peuvent certes échanger des ressources pédagogiques mais ils en perdent le plus souvent les fonctionnalités en changeant d'environnement.

**Mobiliser les chercheurs de l'enseignement des langues comme experts des chantiers de développement des normes**

Quand nous disons que les normes du SC36 sont en cours de développement, nous signifions par là que ce sont des chantiers ouverts dans lesquels les pédagogues de l'apprentissage des langues peuvent encore intervenir. Ceci est un point fondamental. Le fait que ce soit des normes dont la majorité est encore aujourd'hui « en travaux », signifie que les théoriciens de la pédagogie ou spécialistes de l'ingénierie TICE, et pour ce qui nous concerne ceux qui s'intéressent plus précisément à l'enseignement des langues auraient tout intérêt à venir à titre d'expert participer à l'aménagement de toutes ces normes afin d'y préciser les spécifications de fonctionnalités propres au traitement de l'information linguistique et multilinguistique qu'ils aimeraient voir figurer et vérifier si elles sont ou non déjà décrites dans le cadre d'autre fonctionnalités pédagogiques.

Car tel est la méthode de fonctionnement de la construction des normes. A priori les experts en ingénierie pédagogique sont sensés couvrir tous les besoins et dispositifs propres à aménager des normes pour la totalité des besoins propres à tous les styles pédagogiques, tous les modes de médiations (TV, radio, jeux, EAO, e-learning, simulateurs), à toutes les

---

[11] LOM (Learning Objet Metadata)

disciplines… En réalité, la normalisation des TICE se développe par cooptation successives à partir de ses experts pionniers à savoir les militaires et l'aéronautique auxquels sont venus s'adjoindre d'abord les pédagogues des sciences exactes et expérimentales puis les formateurs d'entreprises à forte valeur ajoutée (banque, gestion d'entreprise).

**Décrire puis éventuellement normaliser les fonctionnalités multimédia spécifiques de l'enseignement des langues :**

Le champ des fonctionnalités multimédia déjà normalisées potentiellement spécifiques à l'enseignement des langues est très ouvert. Comme expert des normes des TIC et TICE non spécialistes de l'enseignement des langues, nous pouvons en suggérer certaines. Par exemple, dans la norme MPEG4 l'image est potentiellement générée soit par pixels, soit par synthèse. La synthèse d'image numérique correspondant au positionnement labial (Pandzic et al., 2002 ; Triki-Bchir, 2005) voire même pour le cas particulier d'apprentissage en langue des signes[12] xx celle des mains et autres points du corps peuvent être utilement mis en œuvre dans des plateformes d'enseignement des langues. En MPEG4 on trouve les spécifications FBA (*Face and Body Animation*), les FDP (*Face Definition Parameters*) et les BDP (*Body Definition Parameters*) qui répondent à ces fonctionnalités.

Dans un tout autre ordre d'idée, la mise en parallèle de colonnes de textes et l'alignement de ressources linguistiques écrites dans diverses langues et diverses écritures est très utile. Sont aussi très utiles des normes de translittération, de transcription, les normes permettant l'interopérabilité entre plusieurs écritures ainsi bien sûr que les normes de transcription phonétique en API. En effet dans les systèmes d'EAD d'apprentissage des langues il peut être utile de partir de ressources pédagogiques écrites et de les transformer en ressources audio grâce à deux « narrateurs synthèse vocale[13] » : un dans la langue maternelle de l'apprenant et un autre dans la langue cible de l'apprentissage. Dès lors cela pose la question du développement concerté (et nécessairement normalisé) de « narrateurs synthèse vocale » libres et gratuits en arabe et en berbère cette dernière langue(s) ayant peu et vraisemblablement pas de ressources en la matière. Les outils symétriques de reconnaissance vocale (dictée vocale) sont aussi un enjeu important pour l'apprentissage des langues dans la mesure où ils peuvent permettre à l'élève de contrôler seul qu'il émet correctement un énoncé : l'outil lorsqu'il reconnaît ce qui est dit peut contrôler la conformité phonétique, mais aussi lorsqu'il a reconnu tous les mots la conformité syntaxique de l'énoncé produit et éventuellement sa conformité sémantique (qu'il ne répond pas à coté de la question). Là aussi, comme pour la synthèse vocale, la reconnaissance vocale impose que la communauté des linguistes maghrébins unisse ses efforts dans un cadre normatif consensuel lui permettant de développer si possible des ressources gratuites et libres de droit dans des langues dont la rentabilité est bien plus faible que pour l'anglais.

Les normes lexicologiques et terminologiques (ISO TC 37) sont également très utiles pour être certains de pouvoir réutiliser, augmenter, et surtout associer des formats compatibles de terminologies et de lexiques multilingues. Le format commun (*TMF, Terminological Markup Framework*) correspond à une organisation onomasiologique (Hudrisier & Ben Henda, 2008) (du concept vers les mots dans les langues) modélisée selon un schéma applicatif fondé sur XML. Ces bases de données terminologiques seront non seulement utiles pour apprendre du vocabulaire, contrôler ou assister des exercices de version ou de thème mais elles seront également utiles pour toutes les données sémantiques (définitions) mais aussi morpho-syntaxiques qu'elles contiennent. Elles peuvent rendre d'immenses services tant pour les créateurs de ressources pédagogiques que pour les professeurs, les tuteurs ou plus directement les apprenants.  Il reste, que là encore ces associations en mosaïque de quantités de ressources provenant de sources très diverses ne peuvent fonctionner qu'autant que les créateurs et les éditeurs de ressources connaissent très exhaustivement la structure normative de ces domaines et qu'éventuellement ils auront décrit leurs besoins spécifiques d'enseignants des langues et qu'ils auront demandé aux normalisateurs du multimédia, des terminologies, des TICE d'en tenir compte et éventuellement de développer des contraintes normatives spécialisées.

---

[12] On est là précisément dans un cas de superposition entre normes (1) la liaison MPEG/SC36 est elle suffisamment activée pour s'assurer que l'interopérabilité est possible et sinon qu'il faut la développer (2) on est face à un cas de question normative traitée en principe par le SC36WG7 (handicap) Est-ce traité à fond pour être sûr que ces normes (aussi bien lecture labiale que mouvement du corps des mains et de la face permettront d'assurer une intercomptabilité en direction de la simulation synthèse de langue des signes.

[13] Les « narrateurs synthèse vocale » comme celui de Vista sont des outils logiciels qui permettent de générer des énoncés vocaux à partir de ressources écrites (soit en écriture de langue naturelle, soit à partir de l'API).

Au-delà, tous les experts berbères et arabes doivent unir leurs efforts pour préserver au sein de ces instances de normalisation leurs spécificités nationales linguistiques, culturelles, voire économiques. L'arabe et le berbère sont deux langues (et deux écritures) qui présentent, chacune différemment de nombreuses variantes.

Pour l'arabe il est unifié en tant que langue et en tant qu'écriture dans sa forme classique et canonique. Par contre, les arabes dialectaux qui sont des réalités à prendre en compte ne peuvent être notés que sous des formes accentuées. Les signes diacritiques sont d'une valeur sémantique importante même en arabe littéraire (Ben Henda, 1999). Ils gèrent souvent les ambiguïtés linguistiques inhérentes aux cas d'homonymie[14]. Ces différentes formes linguistiques doivent être nécessairement repérées et codifiées en tant que variantes dans les listes de langues afin que les différentes ressources pédagogiques dans ces différentes formes puissent être distinguées. Cela est d'autant plus nécessaire à l'ère des échanges multimédia (indispensables pour l'apprentissage des langues) qui permettent de disposer de séquences sonores.

Les variantes berbères sont dans une situation encore plus diversifiée qui exigerait que les normalisateurs créent un nombre de codes distincts pour chaque variante de grandes familles de langues (kabyle, mozabite, rifain, touareg, etc..) puis des sous-codes pour les parlers principaux à l'intérieur de ces familles. Ils devraient de plus pouvoir attribuer à chaque ressource un code d'écriture dans les 3 écritures traditionnellement décrites (écriture arabe+lettres additionnelles ; écriture latine+lettres additionnelles ; tifinagh et ses variantes). A cela pour des usages plus spécialisés nous avons évidemment besoin de pouvoir spécifier pour une ressource des écrits en API ou des écrits sous des formes scripturales anciennes (archéologiques).

Au-delà de ces indispensables questions de codification exhaustive des langues et écritures intervient la question de création de terminologies spécialisées en arabe et en berbère tant des TICE et de l'univers d'apprentissage plus précisément de l'apprentissage des langues. Ces terminologies sont indispensables premièrement car il serait contre productif qu'un jeune berbérophone natif doive gérer en français ou même en arabe tout le vocabulaire d'interface (écran, souris, clavier…), d'autre part il en sera de même pour un apprenant japonais ou russe qui devra pouvoir accéder à l'interface d'apprentissage dans sa langue maternelle.

Nous concluons en proposant que les experts magrébins et francophones (tant spécialiste de l'e-Learning, que de la pédagogie ou de la linguistique s'associent en synergie pour décrire et développer un univers normatif multimédia, pédagogique et linguistique apte à défendre leur diversité et leurs spécificités.

---

[14] En arabe non accentué, il est difficile de distinguer la valeur sémantique des deux concepts « Monument » et « Instituteur ». Les exemples sont innombrables en langue arabe littéraire et dialectale.

La norme, malgré sa signification triviale unificatrice et banalisatrice est en réalité le lieu où se défendent au quotidien les spécificités culturelles, linguistiques mais aussi pédagogiques et économiques.

**Bibliographie :**